\documentclass[english,superscriptaddress,showpacs]{revtex4}
\usepackage[T1]{fontenc}
\usepackage[utf8]{inputenc}
\usepackage{graphicx}
\usepackage{amssymb}
\usepackage{babel}

\begin{document}

\title{Shear viscosity in late time of hydrodynamic evolution in AdS/CFT
duality}

\author{Shi Pu}

\affiliation{Institut f\"ur Theoretische Physik, Johann Wolfgang Goethe-Universit\"at,
Max-von-Laue-Str. 1, D-60438, Frankfurt am Main, Germany}

\affiliation{Interdisciplinary Center for Theoretical Studies and Department of
Modern Physics, University of Science and Technology of China, Anhui
230026, China}

\author{Qun Wang}

\affiliation{Interdisciplinary Center for Theoretical Studies and Department of
Modern Physics, University of Science and Technology of China, Anhui
230026, China}

\begin{abstract}
We investigate the shear viscosity $\eta$ and the entropy density
$s$ of strongly coupled $\mathcal{N}=4$ super Yang-Mills (SYM) plasma
in late time of hydrodynamic evolution with AdS/CFT duality and Bjorken
scaling. We use correlation function method proposed by Kovtun, Son
and Starinets. We obtain the metric $g_{\mu\nu}$ in a proper time
dependent $AdS_{5}$ space through holographic renormalization, whose
boundary condition is given by energy-momentum tensor of the plasma
in 2+1 dimension with transverse expansion or radial flow. With the
metric we compute $\eta$ and $s$ of fluids in 1+1 and 2+1 dimension
without and with radial flow. We find the ratio $\eta/s=1/(4\pi)$
in 1+1 dimension consistent with the Kovtun-Son-Starinets bound if
next-to-leading terms in proper time are included in the equation
of motion for metric perturbations. For 2+1 dimension the result is
unchanged in the leading order of transverse rapidity. 
\end{abstract}


\maketitle

\section{Introduction}

Quantum chromodynamics (QCD) tells us that quarks and gluons are confined
inside hadrons in vacuum or ground state. No free quarks and gluons
are detected in any hadronic collisions. Finite temperature theory
of QCD predicts that the deconfinement of quarks and gluons can be
reached at high temperatures and densities and a new state of matter,
the quark gluon plasma (QGP), can be created \citep{Lee:1974ma}.
According to lattice QCD calculations the confinement-deconfinement
phase transition takes place at temperature of about 170 MeV in three
flavor case \citep{Karsch:2000ps}. In early 1970s Lee and Greiner
and his collaborators proposed that the deconfinement could be realized
by smashing two heavy ions together at ultrarelativistic energies
\citep{Hofmann:1974}. This is the only way of generating QGP in laboratories.
The minimum energy density for deconfinement to take place is about
an order of magnitude higher than normal nuclear matter density. The
current energy frontier of heavy ion collisions is the Relativistic
Heavy Ion Collider (RHIC) at Brookhaven National Laboratory (BNL),
which has been running since the summer of 2000. A lot of evidences
imply that the QGP at RHIC is near perfect fluid (strongly coupled
QGP or sQGP), which is contrary to the conventional concept of QGP
as a weakly interacting gas of quarks and gluons, see e.g. \citet{Shuryak:2004cy,gyulassy:2005}.

While perturbation can be used to deal with weak coupling problems,
there are a lot of difficulties in describing a strongly coupled system
due to its non-linear feature and failure of perturbative methods.
In recent years a promising method to deal with strong coupling problems
in gauge theory is through string/gauge duality or precisely AdS/CFT
duality, where AdS and CFT are abbreviations for anti-de Sitter space
and conformal field theory respectively. The duality was first proposed
by Maldacena and many others \citep{Maldacena:1997re,Gubser:1998bc,Witten:1998zw},
where a strongly coupled conformal gauge field theory corresponds
to weekly coupled closed strings at 't Hooft limit. So properties
of a weekly coupled string can provide information for strongly coupled
gauge theory. 

Policastro, Son and Starinets first used AdS/CFT duality to calculate
transport coefficients of strongly coupled dense matter \citep{Policastro:2001yc}.
Kovtun, and Son and Starinets derived a lower bound (called KSS bound)
for the ratio of shear viscosity to entropy density, $\eta/s\gtrsim1/(4\pi)$,
consistent to the RHIC data \citet{Kovtun:2004de}. These works open
an avenue to study properties of strongly coupled matter followed
by a lot of developments in hydrodynamic properties of sQGP \citet{Buchel:2003tz,Maeda:2006by,Natsuume:2007ty,Baier:2007ix}.
Recently the jet quenching effect has been investigated with AdS/CFT
techniques \citet{Liu:2006ug,Herzog:2006gh,Gubser:2006bz}. The sceening
length of the heavy quark potential can also be described from the
AdS/CFT duality \citet{Liu:2006nn,Peeters:2006iu,Hou:2007uk,Li:2008py},
which is closely related to the $J/\psi$ suppression in hot medium. 

The KSS bound is valid for any conformal fluids. Altough it is debatable
that the sQGP is a conformal fluid, the ratio $\eta/s$ is fixed to
be the KSS bound in all periods of the sQGP evolution in some hydrodynamic
simulations \citep{Song:2007ux,Chaudhuri:2008sj}. However the sQGP
formed in ultra-relativistic heavy ion collisions has particular geometric
configurations such as boost invariance \citep{Bjorken:1982qr} etc..
Janik and Peschanski first considered a proper time dependent $AdS_{5}$
metric with boost invariance through holographic renormalization in
late time of fluid evolution \citet{Janik:2005zt,Janik:2006ft,Benincasa:2007tp}.
The method has been extended to sQGP with shear viscosity \citep{Nakamura:2006ih,Sin:2006pv},
where the shear viscosity appears in energy-momentum tensor in CFT
and finally enter in the $AdS_{5}$ metric. The early time behavior
of sQGP has also been addressed following similar method and can then
join the late time one to provide a global picture for time evolution
in heavy ion collision \citet{Kovchegov:2007pq,Albacete:2008vs}.
The exact solution to the gravity dual of 1+1 dimensional sQGP with
Bjorken scaling has been found by Kajantie, Louko and Tahkokallio,
which is helpful to understand what happens in 1+4 dimension in the
real world \citet{Kajantie:2007bn}. The proper time dependent $AdS_{5}$
metric without shear viscosity does not necessarily mean that the
${\cal N}=4$ super Yang-Mills (SYM) plasma is non-viscous, noting
that the shear viscosity with AdS/CFT duality is obtained by a D3
black AdS metric in Ref. \citep{Son:2002sd,Son:2007vk}. If we treat
the proper time dependent $AdS_{5}$ metric without viscosity as an
extension of standard AdS metric we can build up a connection between
the sQGP as an ideal fluid to the SYM plasma, then the ratio $\eta/s$
may not be zero. 

In this paper, we investigate the 2+1-dimensional expansion (also
including 1+1-dimensional expansion) of the sQGP in late time and
obtain the proper time dependent $AdS_{5}$ metric through holographic
renormalization. The metric dual to the sQGP fluid can be obtained
by solving the Einstein equation with the stress tensor of sQGP as
boundary value of the metric if transverse expansion is small and
can be treated as a perturbation. The metric has off-diagonal elements
proportional to transverse rapidity. We compute with the metric the
shear viscosity in 1+1 and 2+1 dimension through Kubo formula with
AdS/CFT duality. In 1+1 dimension with transverse expansion turned
off, the shear viscosity is found to be vanishing if only the leading
terms in proper time are considered in the Einstein equation or the
equation of motion for perturbations to the metric. Only including
the next-to-leading terms can one get the ratio $\eta/s=1/(4\pi)$,
indicating the shear viscosity is a higher order effect in late time
behavior. In 2+1 dimension the result is unchanged in the leading
order of transverse rapidity. 

The structure of the paper is as follows. In Sec. \ref{sec:AdS5},
we obtain the proper time dependent $AdS_{5}$ metric corresponding
to the sQGP in 2+1 dimension. In Sec. \ref{sec:Boost-geometry} we
derive the effective action for perturbations in the AdS metric. We
compute the ratio $\eta/s$ in Sec. \ref{sec:ratio1+1} and \ref{sec:ratio2+1}
for 1+1 and 2+1 dimensional cases respectively. We make summary and
conclusions in Sec. \ref{sec:Conclusion}. In the paper we take the
signature $(-,+,+,+,+)$ for the metric.

\section{2+1 hydrodynamics with proper time dependent $AdS_{5}$ metric\label{sec:AdS5}}

\subsection{2+1 hydrodynamics for ideal fluids}

In order to describe the hydrodynamic evolution of an ultrarelativistic
system with transverse expansion, e.g. in heavy ion collisions, we
use cylindrical coordinates, \begin{eqnarray}
x^{\mu} & = & (\tau,y,r,\theta),\label{eq:dim4}\end{eqnarray}
where $y$ is the rapidity, $\tau$ the transverse proper time, $r$
the radius in transverse plane, and $\theta$ the azimuthal angle
around longitudinal direction. Note that the full proper time $\tau_{f}$
is related to $\tau$ by $\tau_{f}^{2}=\tau^{2}-r^{2}$. Hereafter
the proper time means the transverse one if not explicitly stated.
The line element is given by \begin{equation}
ds^{2}=-d\tau^{2}+\tau^{2}dy^{2}+dr^{2}+r^{2}d\theta^{2},\label{eq:metric-dim4}\end{equation}
which gives the metric $\widetilde{g}_{\mu\nu}^{(0)}=\mathrm{diag}(-1,\tau^{2},1,r^{2})$.
The velocity four-vector can be parametrized as \begin{eqnarray}
u^{\mu} & = & \frac{dX^{\mu}}{d\tau_{f}}=\frac{1}{\sqrt{1-R^{2}(\tau,r)/\tau^{2}}}\left[1,0,\frac{R(\tau,r)}{\tau},0\right]\nonumber \\
 & \equiv & \left[\cosh\alpha(\tau,r),0,\sinh\alpha(\tau,r),0\right],\end{eqnarray}
where $R$ is a function of $(\tau,r)$ and $\alpha$ is the radial
rapidity in transverse direction. The radial velocity is defined by
$u_{r}\equiv\sinh\alpha$. One can verify that the velocity satisfies
$u^{\mu}u_{\mu}=-1$.

\begin{figure}[t]
\includegraphics[scale=0.42]{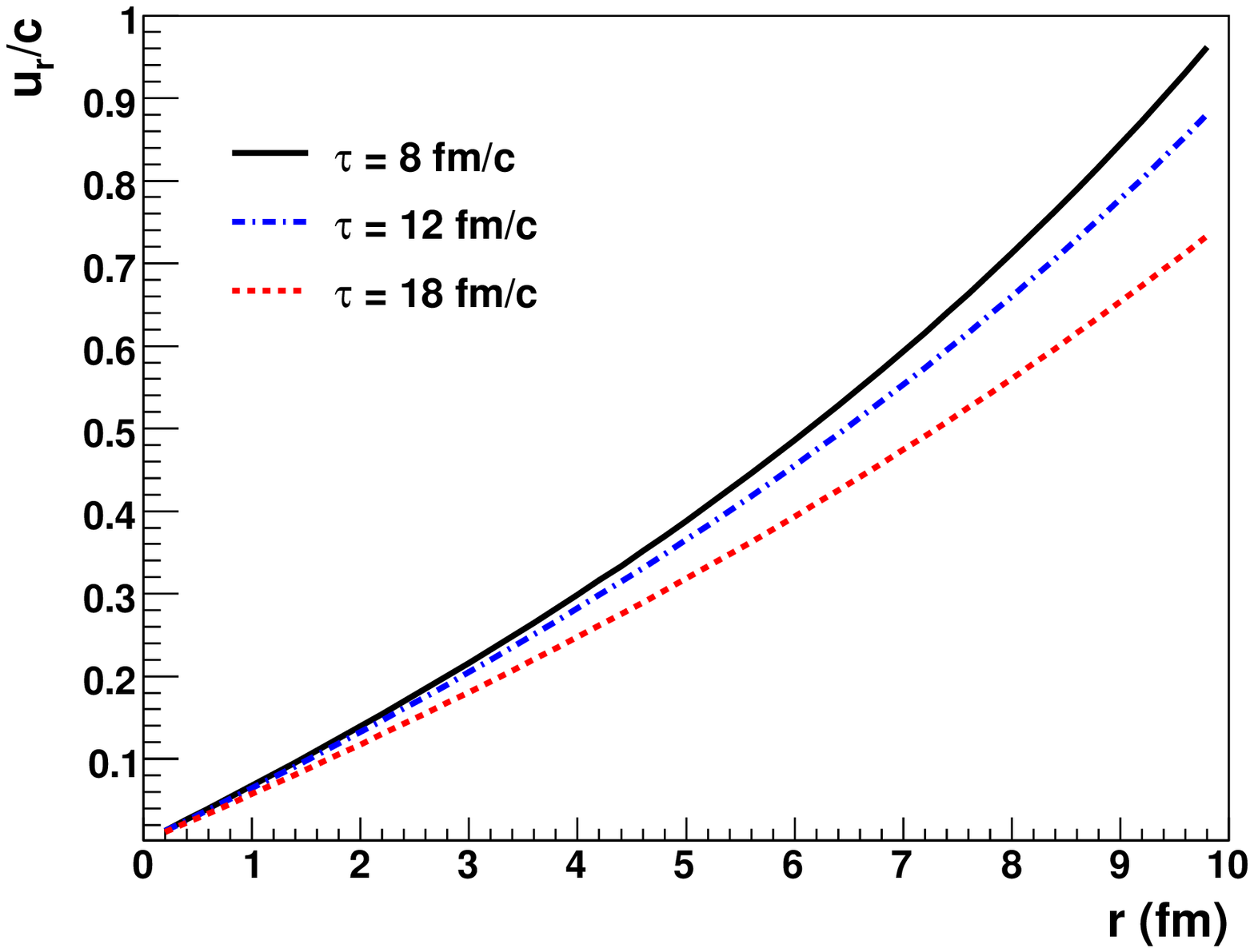} \includegraphics[scale=0.42]{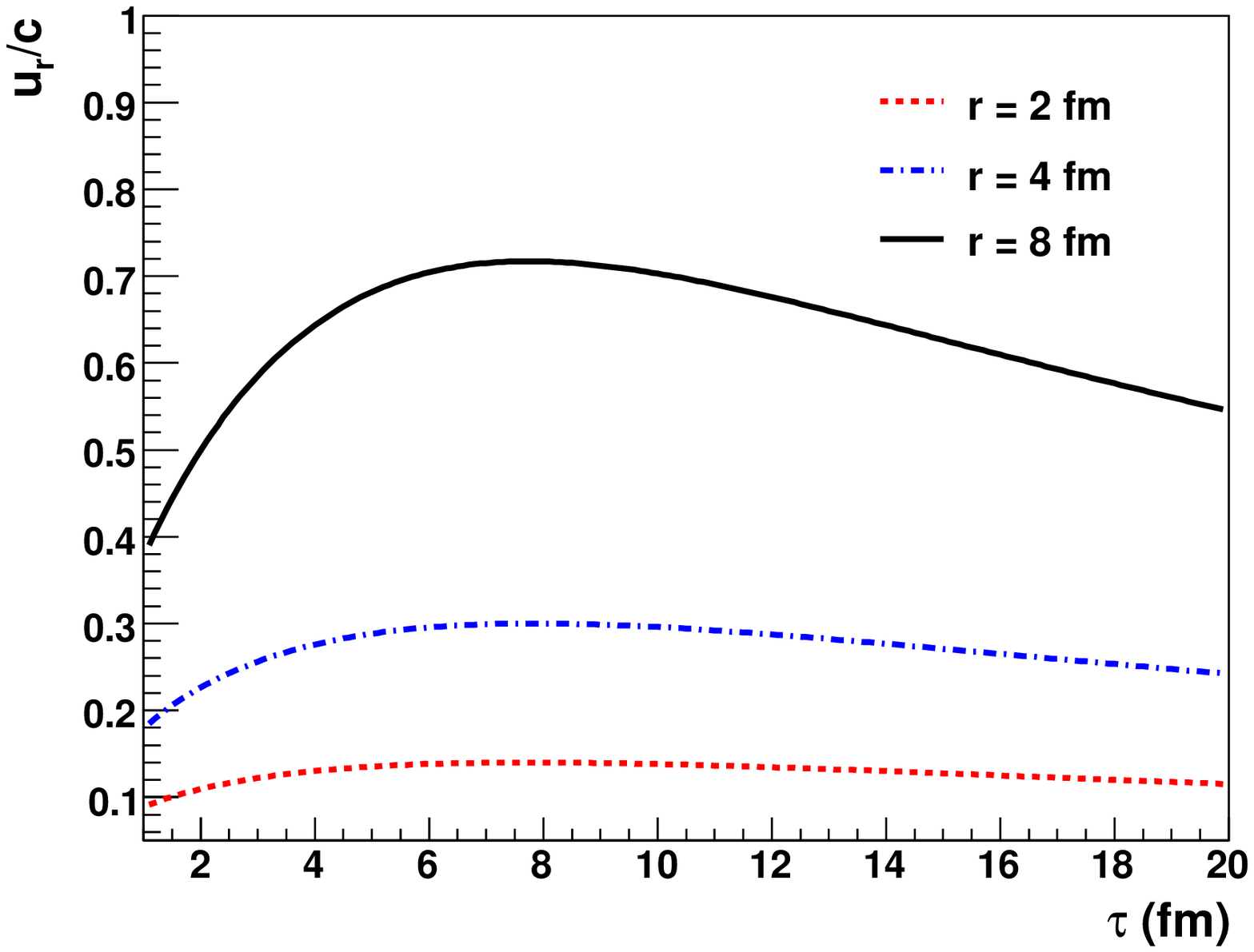} 

\caption{\label{fig:radial-velocity}{\small The radial velocity $u_{r}=\sinh\alpha$
as functions of $r$ and $\tau$.}}

\end{figure}

\begin{figure}
\includegraphics[scale=0.45]{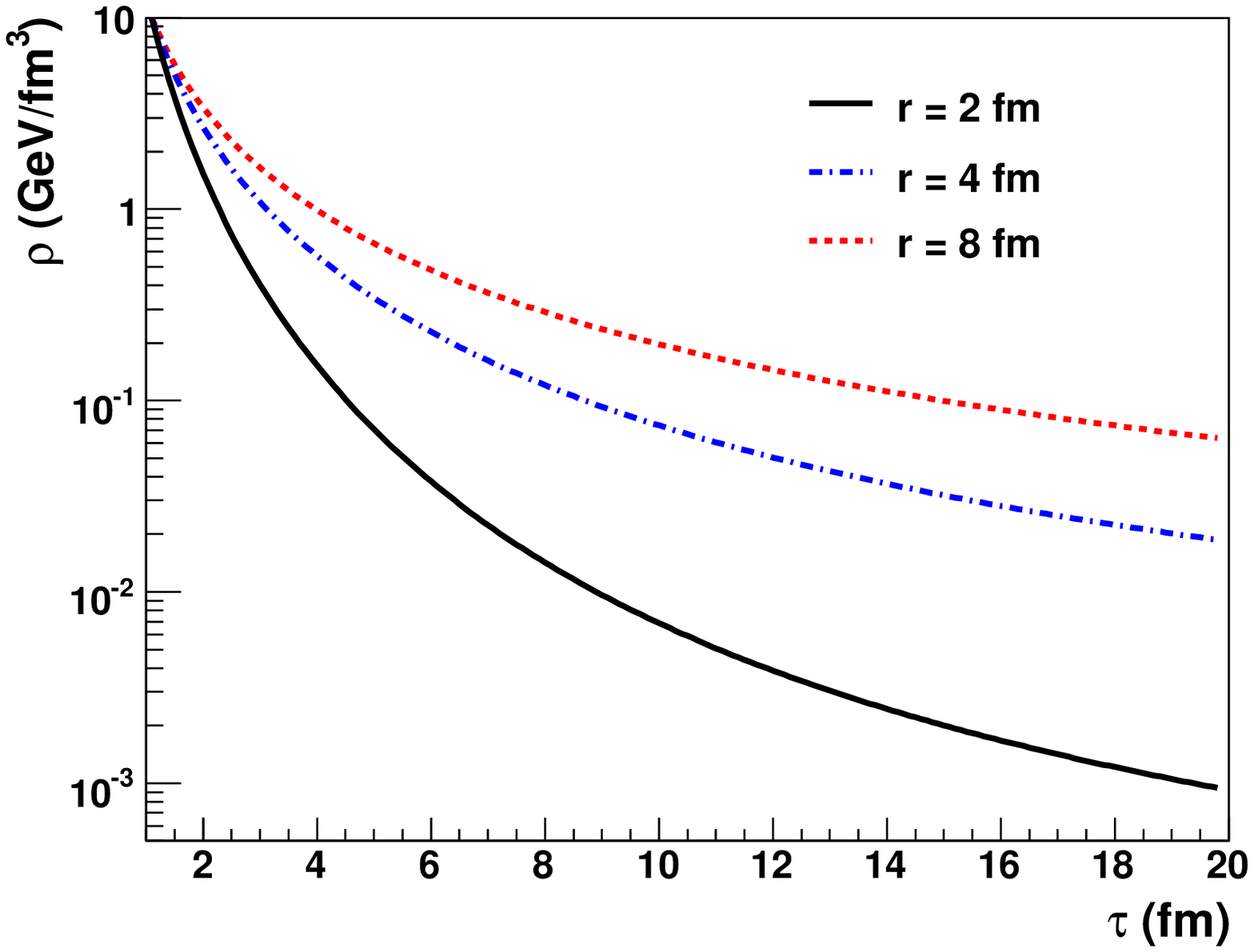}

\caption{\label{fig:density-time}{\small The radial velocity $u_{r}=\sinh\alpha$
as functions of $r$ and $\tau$.}}

\end{figure}

For ideal fluid with the equation of state $\rho=3P$ where $\rho$
and $P$ are energy density and pressure respectively, the energy-momentum
tensor is \begin{eqnarray}
T_{\mu\nu} & = & \left(\rho+P\right)u_{\mu}u_{\mu}+P\widetilde{g}_{\mu\nu}^{(0)}(x)\nonumber \\
 & = & \left(\begin{array}{cccc}
-\frac{\rho}{3}+\frac{4}{3}\rho\cosh^{2}\alpha & 0 & -\frac{4}{3}\rho\sinh\alpha\cosh\alpha & 0\\
0 & \frac{\rho}{3}\tau^{2} & 0 & 0\\
-\frac{4}{3}\rho\sinh\alpha\cosh\alpha & 0 & \frac{\rho}{3}+\frac{4}{3}\rho\sinh^{2}\alpha & 0\\
0 & 0 & 0 & \frac{\rho}{3}r^{2}\end{array}\right).\label{eq:EMT01}\end{eqnarray}
One can verify that $T_{\mu\nu}$ satisfies conformal invariance $T_{\mu}^{\mu}=0$.
The energy-momentum conservation equations read \begin{equation}
\nabla_{\mu}T^{\mu\nu}\equiv\partial_{\mu}T^{\mu\nu}+\Gamma_{\mu\sigma}^{\mu}T^{\sigma\nu}+\Gamma_{\mu\sigma}^{\nu}T^{\mu\sigma}=0,\end{equation}
where $\Gamma_{\mu\sigma}^{\mu}$ are Christoffel symbols for the
metric in (\ref{eq:metric-dim4}). The above equations lead to \begin{eqnarray}
\partial_{\tau}\ln\rho & = & -\frac{2}{\tau}\frac{2\cosh^{2}\alpha}{2\cosh^{2}\alpha+1}\left(1+\frac{R}{\tau}+\partial_{\tau}R\right),\nonumber \\
\partial_{\tau}\ln R & = & \frac{1}{\tau}\frac{R/\tau+2+2(1-\partial_{r}R)\cosh^{2}\alpha}{2\cosh^{2}\alpha+1}.\label{eq:evolution}\end{eqnarray}
Given the initial condition $R(\tau_{0},r)=\xi\tau_{0}r$ where $\xi$
is set to 0.05 $\mathrm{fm}^{-1}$, Eq. (\ref{eq:evolution}) can
be solved numerically whose results are shown in Fig. (\ref{fig:radial-velocity})(\ref{fig:density-time})
and are comparable with those in Ref. \citep{Kolb:2003dz}. The results
show that the radial velocity is linearly proportional to the radial
distance, i.e. $u_{r}\sim\xi r$, where $\xi$ is a small number.
One can see in the figure that $u_{r}$ rises sharply in a very short
early time and gradually falls with increasing $\tau$. If the expansion
in transverse direction is negligible compared to that in longitudinal
direction the energy density damps in the same way as in the case
of 1+1 dimension, i.e. $\rho\sim\tau^{-4/3}$.

\subsection{General setup of holographic model for 2+1 hydrodynamics\label{sub:General-setup-of}}

In this section we briefly introduce the general idea of the holographic
model for relativistic hydrodynamics. The metric in $AdS$ space can
in general be written as \citep{fefferman1985}, \begin{eqnarray}
ds^{2} & = & g_{MN}dX^{M}dX^{N}=\frac{1}{z^{2}}\left[\widetilde{g}_{\mu\nu}(x,z)dx^{\mu}dx^{\nu}+dz^{2}\right],\label{eq:gen-metric01}\end{eqnarray}
where $X^{M}=(x^{\mu},z)$ and $\widetilde{g}_{\mu\nu}(x,z)$ is the
metric tensor in the 4-dimensional space (\ref{eq:dim4}) and generally
depends on $z$. Normally one assumes the following form of $\widetilde{g}_{\mu\nu}(x,z)$,
\begin{equation}
\widetilde{g}_{\mu\nu}(x,z)dx^{\mu}dx^{\nu}=-Ad\tau^{2}+B\tau^{2}dy^{2}+C\left(dr^{2}+r^{2}d\theta^{2}\right)+2Dd\tau dr,\end{equation}
where we have introduced the off-diagonal term proportional to $d\tau dr$
to account for transverse expansion or radial flow implied by energy-momentum
tensor in (\ref{eq:EMT01}). Note that the off-diagonal term related
to transverse expansion is absent in the metric (\ref{eq:metric-dim4}).
One can write the coefficients $A$, $B$, $C$ and $D$ in exponential
forms, \begin{equation}
A=e^{w},\; B=e^{b},\; C=e^{c},\; D=e^{d}.\end{equation}
In general these coefficients are functions of $x$ and $z$. A standard
way of holographic renormalization \citep{Karch:2005ms,Skenderis:2002wp}
is to expand the metric $\widetilde{g}_{\mu\nu}(x,z)$ or these coefficients
in the fifth coordinate $z$, \begin{equation}
\widetilde{g}_{\mu\nu}(x,z)=\sum_{n=0}z^{2n}\widetilde{g}_{\mu\nu}^{(2n)}(x),\end{equation}
where $\widetilde{g}_{\mu\nu}^{(0)}(x)$ is just the metric (\ref{eq:metric-dim4})
in 4-dimensional space. The second order term $\widetilde{g}_{\mu\nu}^{(2)}(x)$
can be proved to be vanishing. The fourth order term $\widetilde{g}_{\mu\nu}^{(4)}(x)$
is given by the energy-momentum tensor, \begin{equation}
\widetilde{g}_{\mu\nu}^{(4)}(x)\propto\left\langle T_{\mu\nu}\right\rangle ,\end{equation}
where we set factor $\frac{N^{2}}{2\pi^{2}}=1$ ($N$ is the number
of colors) and we will finally restore it. Given by these boundary
conditions one can solve the $AdS$ metric (\ref{eq:gen-metric01})
from the Einstein equation with the cosmological constant $\Lambda=-6$
which we will discuss about in Sec. (\ref{sec:Boost-geometry}), \begin{equation}
R_{MN}-\frac{1}{2}g_{MN}R+6g_{MN}=0,\label{eq:einstein01}\end{equation}
where $R_{MN}$ and $R$ are the curvature tensor and scalar in $AdS$
space with the metric $g_{MN}$ given in (\ref{eq:gen-metric01}).

\subsection{Holographic model in late time}

The introduction of the $r$ dependence of the metric (\ref{eq:metric01})
makes solving the Einstein equation a formidable task even with the
method of Janik and Peschanski \citep{Janik:2005zt,Nakamura:2006ih,Sin:2006pv}.
So the simplification is necessary. In this section we consider a
simplified problem with perturbation in transverse direction based
on the Janik-Peschanski method. 

Supposing $\alpha$ is small in Eq. (\ref{eq:EMT01}) we can make
an expansion of $T_{\mu\nu}$ in $\alpha$,

\begin{eqnarray}
T_{\mu\nu} & = & \left(\begin{array}{cccc}
\rho & 0 & 0 & 0\\
0 & \tau^{2}\frac{1}{3}\rho & 0 & 0\\
0 & 0 & \frac{1}{3}\rho & 0\\
0 & 0 & 0 & r^{2}\frac{1}{3}\rho\end{array}\right)+\alpha\left(\begin{array}{cccc}
0 & 0 & -\frac{4}{3}\rho & 0\\
0 & 0 & 0 & 0\\
-\frac{4}{3}\rho & 0 & 0 & 0\\
0 & 0 & 0 & 0\end{array}\right)\nonumber \\
 &  & +\alpha^{2}\left(\begin{array}{cccc}
\frac{4}{3}\rho & 0 & 0 & 0\\
0 & 0 & 0 & 0\\
0 & 0 & \frac{4}{3}\rho & 0\\
0 & 0 & 0 & 0\end{array}\right)+O\left(\alpha^{3}\right).\label{eq:expansion01}\end{eqnarray}
In this expansion, we see that off-diagonal elements are proportional
to odd power of $\alpha$, while modifications of diagonal elements
are proportional to even power of $\alpha$. In leading order the
transverse velocity vanishes and we reproduce the results in Ref.
\citep{Janik:2005zt,Nakamura:2006ih}. In next-to-leading order $O(\alpha)$
the transverse velocity does not vanish while the energy density and
pressure do not change, whose modifications are in $O(\alpha^{2})$
or higher. 

If we assume that there is no linear terms of $\alpha$ in diagonal
components of the AdS metric $g_{MN}$, we get the following result
by solving the Einstein equation (\ref{eq:einstein01}), \begin{equation}
g_{MN}=\frac{1}{z^{2}}\left(\begin{array}{ccccc}
-\frac{(1-a)^{2}}{1+a} & 0 & \frac{4}{3}\alpha\frac{(1-a)^{2}}{1+a} & 0 & 0\\
0 & \tau^{2}(1+a) & 0 & 0 & 0\\
\frac{4}{3}\alpha\frac{(1-a)^{2}}{1+a} & 0 & 1+a & 0 & 0\\
0 & 0 & 0 & r^{2}(1+a) & 0\\
0 & 0 & 0 & 0 & 1\end{array}\right),\label{eq:solution01}\end{equation}
with $a=\frac{\rho_{0}z^{4}}{3\tau^{4/3}}$ and $\rho_{0}$ is a constant
energy density. 

In the case without off-diagonal components in the metric (\ref{eq:solution01}),
i.e. $\alpha=0$, the line element in central rapidity region (not
for non-central rapidity region) has the form, \begin{equation}
ds^{2}=-\frac{(1-z^{4}/z_{H}^{4})^{2}}{1+z^{4}/z_{H}^{4}}\frac{d\tau^{2}}{z^{2}}+\frac{1+z^{4}/z_{H}^{4}}{z^{2}}[\tau^{2}dy^{2}+dr^{2}+r^{2}d\theta^{2}]+\frac{dz^{2}}{z^{2}},\label{eq:static-metric}\end{equation}
where the black hole locates at \begin{equation}
z_{H}=\left(\frac{\rho_{0}}{3}\right)^{-1/4}\tau^{1/3},\label{eq:bh-zh}\end{equation}
which depends on $\tau$. See Appendix (\ref{sec:app-b}) for the
derivation of the metric (\ref{eq:static-metric}). Changing variables,
\begin{equation}
z\rightarrow\widetilde{z}=\frac{z}{\sqrt{1+\frac{z^{4}}{z_{H}^{4}}}},\qquad z_{H}\rightarrow\widetilde{z}_{H}=\frac{z_{H}}{\sqrt{2}},\label{eq:trans01}\end{equation}
we can verify that the line element (\ref{eq:static-metric}) is the
standard one in D3 black $AdS_{5}$ space if $z_{0}$ is constant,
see Sec. \ref{sec:Boost-geometry}, \begin{equation}
ds^{2}=-\frac{1-\widetilde{z}^{4}/\widetilde{z}_{H}^{4}}{\widetilde{z}^{2}}d\tau^{2}+\frac{d\mathbf{x}^{2}}{\widetilde{z}^{2}}+\frac{1}{1-\widetilde{z}^{4}/\widetilde{z}_{H}^{4}}\frac{d\widetilde{z}^{2}}{\widetilde{z}^{2}},\label{eq:ads-metric01}\end{equation}
where $d\mathbf{x}^{2}\equiv\tau^{2}dy^{2}+dr^{2}+r^{2}d\theta^{2}$
and $\widetilde{z}_{H}$ is the location of the black hole horizon,
the cosmology constant for $AdS_{d+1}$ is \citep{Skenderis:2002wp}
\begin{equation}
\Lambda=-\frac{d(d-1)}{2}=-6,\label{eq:cosmolog}\end{equation}
and the Hawking temperature is determined by the behavior of the metric
near the horizon, \begin{equation}
T_{H}=\frac{1}{\pi\widetilde{z}_{H}}=\frac{\sqrt{2}}{\pi z_{H}}.\label{Hawking temp}\end{equation}

Hereafter we call the proper time dependent $AdS_{5}$ metric the
$AdS^{\#}$ one in (\ref{eq:solution01}) with transverse expansion
or in (\ref{eq:static-metric}) without transverse expansion.

\section{Effective action for perturbations in $AdS$ metric\label{sec:Boost-geometry}}

The low energy limit of type IIB string theory in $AdS_{5}\times S^{5}$
space can be approximated by a five-dimensional theory (see, e.g.
\citet{Son:2007vk}) whose action is \begin{equation}
I_{5D}\approx\frac{N^{2}}{8\pi^{2}R^{3}}\int d^{5}x(\mathcal{R}_{5D}-2\Lambda+\cdots),\label{eq:action01}\end{equation}
where $\mathcal{R}_{5D}$ is the scalar curvature in $AdS_{5}$, $\Lambda$
the cosmological constant, and $R$ the radius of $S^{5}$. The line
element in $AdS_{5}$ space without black hole is \begin{equation}
ds^{2}=\frac{r^{2}}{R^{2}}(-dt^{2}+d\mathbf{x}^{2})+\frac{R^{2}}{r^{2}}dr^{2}.\label{eq:metric01}\end{equation}
Using the variable $z=R^{2}/r$, the above line element can be written
as \begin{equation}
ds^{2}=\frac{R^{2}}{z^{2}}(-dt^{2}+d\mathbf{x}^{2}+dz^{2}).\label{eq:metric02}\end{equation}
The line element with a black hole has the form \begin{eqnarray}
ds^{2} & = & \frac{R^{2}}{z^{2}}\left[-(1-z^{4}/z_{H}^{4})dt^{2}+d\mathbf{x}^{2}+\frac{1}{1-z^{4}/z_{H}^{4}}dz^{2}\right]\nonumber \\
 & = & \frac{\pi^{2}T^{2}R^{2}}{u}[-(1-u^{2})dt^{2}+d\mathbf{x}^{2}]+\frac{R^{2}}{4u^{2}(1-u^{2})}du^{2},\label{eq:metric03}\end{eqnarray}
where $u=z^{2}/z_{H}^{2}$ and $z_{H}=1/(\pi T)$. 

The AdS/CFT correspondence can be expressed by \begin{equation}
Z_{4D}[J]=\left.e^{iS_{5D}[\phi]}\right|_{J(x)=\phi(x,z=0)},\label{eq:duality}\end{equation}
where $Z_{4D}[J]$ is the partition function of CFT in 4-dimensional
Minkowski space and $S_{5D}[\phi]$ the action in $AdS_{5}$ given
by (\ref{eq:action01}) for the classical bulk field $\phi(x,z)$,
$J(x)$ is the source in 4-dimension taking the value of the bulk
field on the boundary $z=0$ in $AdS_{5}$. The partition function
$Z_{4D}[J]$ is a functional of $J(x)$, \begin{equation}
Z_{4D}[J]=\int[d\varphi]\mathrm{exp}\left\{ iS_{CFT}[\varphi]+i\int d^{4}xJO[\varphi]\right\} ,\end{equation}
where $\varphi$ denotes the CFT fields, $O[\varphi]$ is the operator
coupling to $J(x)$, and $S_{CFT}[\varphi]$ is the CFT action. Following
the AdS/CFT correspondence the Green functions of operators $O$ in
CFT can be evaluated in terms of $S_{5D}[\phi]$ in $AdS_{5}$, for
example, a two-point Green function is \begin{equation}
\left\langle TO(x_{1})O(x_{2})\right\rangle _{4D}=\left.\frac{\delta^{2}S_{5D}[\phi]}{i\delta\phi(x_{1})\delta\phi(x_{2})}\right|_{J(x)=\phi(x,z=0)}.\label{eq:green-f}\end{equation}

We note that the stress tensor is coupled with the metric, \begin{equation}
T_{\mu\nu}=-\frac{2}{\sqrt{-\widetilde{g}}}\frac{\delta S_{CFT}}{\delta\widetilde{g}_{\mu\nu}},\end{equation}
where $T_{\mu\nu}$ can be regarded as an operator in CFT and the
metric $\widetilde{g}_{\mu\nu}$ in 4-dimension as its source. In
order to obtain the shear viscosity $\eta$ through the Kubo formula,
\begin{equation}
\eta=-\lim_{\omega\rightarrow0}\frac{1}{\omega}\mathsf{\mathrm{Im}}G_{12,12}^{R}(\omega,\mathbf{0}),\label{eq:kubo01}\end{equation}
we need to know the Green function of $T_{12}$ at low energy limit,
\begin{equation}
G_{12,12}^{R}(\omega,\mathbf{k})=-i\int d^{4}xe^{ik\cdot x}\Theta(t)\left\langle [T_{12}(x),T_{12}(0)]\right\rangle .\label{eq:kubo02}\end{equation}
To obtain $G_{12,12}^{R}(\omega,\mathbf{k})$ from the gravitational
dual we should know the action $S_{5D}[g_{\mu\nu}]$ in $AdS_{5}$
as a functional of $g_{\mu\nu}$ with the boundary value $g_{\mu\nu}(z=0)\rightarrow\widetilde{g}_{\mu\nu}$.
For this purpose we consider a variation in the $AdS_{5}$ metric,
$g_{\mu\nu}=g_{\mu\nu}^{(0)}+h_{\mu\nu}$. We can choose for simplicity
$h_{z\mu}=h_{\mu z}=0$ and that $h_{\mu\nu}$ depend on $t$, $x^{3}$
and $z$ (or $u$). The line element in (\ref{eq:metric03}) then
becomes \begin{equation}
ds^{2}=\frac{\pi^{2}T^{2}R^{2}}{u}[-(1-u^{2})dt^{2}+d\mathbf{x}^{2}]+\frac{R^{2}}{4u^{2}(1-u^{2})}du^{2}+h_{\mu\nu}dx^{\mu}dx^{\nu}.\label{eq:metric04}\end{equation}
We can expand the action (\ref{eq:action01}) in terms of $h_{\mu\nu}$,
\begin{eqnarray}
I_{5D} & \approx & \frac{N^{2}}{8\pi^{2}}\int d^{5}x\sqrt{-g}(\mathcal{R}_{\mu\nu}-\frac{1}{2}g_{\mu\nu}\mathcal{R}+\Lambda g_{\mu\nu})h^{\mu\nu}\nonumber \\
 & = & \frac{N^{2}}{8\pi^{2}}\int d^{5}x\sqrt{-g}(\mathcal{R}_{\mu\nu}^{(0)}-\frac{1}{2}g_{\mu\nu}^{(0)}\mathcal{R}_{(0)}+\Lambda g_{\mu\nu}^{(0)})h^{\mu\nu}\nonumber \\
 &  & +\frac{N^{2}}{8\pi^{2}}\int d^{5}x\sqrt{-g}(\delta\mathcal{R}_{\mu\nu}-\frac{1}{2}h_{\mu\nu}\mathcal{R}_{(0)}-\frac{1}{2}g_{\mu\nu}^{(0)}\delta\mathcal{R}+\Lambda h_{\mu\nu})h^{\mu\nu}\nonumber \\
 & = & \frac{N^{2}}{8\pi^{2}}\int d^{5}x\sqrt{-g}(\delta\mathcal{R}_{\mu\nu}-\frac{1}{2}g_{\mu\nu}^{(0)}\delta\mathcal{R}+4h_{\mu\nu})h^{\mu\nu},\label{eq:ation02}\end{eqnarray}
where we have used the fact that the first term of the second equality
is vanishing from the Einstein equation for the unperturbed metric
$g_{\mu\nu}^{(0)}$. We have also used $\mathcal{R}_{0}=-20$ and
$\Lambda=-6$. Here we have chosen the $S^{5}$ radius $R=1$. We
consider the components $h_{12}=h_{21}$ which can be verified to
decouple from others. For the metric (\ref{eq:metric04}), the action
(\ref{eq:ation02}) can be simplified as \begin{eqnarray}
I_{5D} & \approx & \frac{N^{2}}{8\pi^{2}}\int d^{5}x\sqrt{-g}(\delta R_{12}+4h_{12})h^{12}=\frac{N^{2}}{8\pi^{2}}\int d^{5}x\sqrt{-g}\left(-\frac{1}{2}g^{\mu\nu}\partial_{\mu}\phi\partial_{\nu}\phi+\cdots\right),\label{eq:action03}\end{eqnarray}
with $\phi=h_{2}^{1}=g^{11}h_{12}$ and $h^{12}=-g^{11}g^{22}h_{12}$.
The retarded two-point Green functions of the CFT stress tensor can
then be derived from the quadratic terms in $I_{5D}$ as above by
using Eq. (\ref{eq:green-f}). The shear viscosity is then obtained
through the Kubo formula (\ref{eq:kubo01}).

\section{Shear viscosity and entropy density in $AdS_{5}^{\#}$ }

In this section we will calculate the shear viscosity in $AdS_{5}^{\#}$
following the procedure of AdS/CFT duality presented in the previous
section. Also we will compute the entropy density in $AdS_{5}^{\#}$.
We consider fluid evolutions in 1+1/2+1 dimension without/with transverse
expansion or radial flow.

\subsection{1+1 dimension without transverse expansion\label{sec:ratio1+1} }

Following Janik and Peschanski's solution \citep{Janik:2005zt} in
late time, the line element in $AdS_{5}^{\#}$ in 1+1 dimension can
be written as\begin{equation}
ds^{2}=\frac{1}{z^{2}}\left\{ -\frac{(1-a)^{2}}{1+a}d\tau^{2}+(1+a)[\tau^{2}dy^{2}+dx_{1}^{2}+dx_{2}^{2}]\right\} +\frac{dz^{2}}{z^{2}},\label{eq:dim11-metric}\end{equation}
where $a=\frac{\rho_{0}z^{4}}{3\tau^{4/3}}$. Note that the derivative
of $a$ is non-zero, i.e. $\partial_{\tau,z}a\neq0$ in computing
the Ricci tensor and the scalar curvature. Finally we will keep a
constant $a$ while taking the limit $\tau\rightarrow\infty$ when
doing power counting in $\tau$. In adopting the above measure we
can verify after a lenghy but straightforward algebra that the Ricci
tensor $\mathcal{R}_{\mu\nu}=-4g_{\mu\nu}$ and $\mathcal{R}_{5D}=-20$
upto higher order terms in negative powers of $\tau$. The determinant
of the metric is $\sqrt{-g}=\frac{1}{z^{5}}(1-a^{2})\tau$. Taking
a perturbation in the metric $g_{\mu\nu}=g_{\mu\nu}^{(0)}+h_{\mu\nu}$
with $h_{12}=h_{12}(\tau,y,z)$ and $h_{\mu\nu}=0$ for all other
indices $\mu,\nu$, we obtain the quadratic terms of the effective
action up to higher order contributions in negative powers of $\tau$,
\begin{equation}
I_{5D}\approx-\frac{N^{2}}{16\pi^{2}}\int d^{5}x\sqrt{-g}g^{\mu\nu}\partial_{\mu}\phi\partial_{\nu}\phi,\label{eq:action01-1}\end{equation}
where $\phi\equiv h_{2}^{1}=g^{11}h_{12}$, $d^{5}x=d\tau dydx_{1}dx_{2}dz$
and $\mu=\tau,y,1,2,z$. The equation of motion for $\phi$ is then
\begin{equation}
\partial_{\mu}(\sqrt{-g}g^{\mu\nu}\partial_{\nu}\phi)=0.\label{eq:EOM001}\end{equation}
We assume a factorized form for $\phi$, \begin{equation}
\phi(\tau,y,z)=\phi_{0}(y,\tau)f(z)=\int\frac{d\omega dp_{3}}{(2\pi)^{2}}\exp(-i\omega\tau\cosh y+ip_{3}\tau\sinh y)\phi_{0}(\omega,p_{3})f_{p}(z)\label{eq:factorization}\end{equation}
where $p_{3}$ is the momentum along the third axis (we do not distinguish
subscript or superscript for this mometum). Inserting the above into
Eq. (\ref{eq:EOM001}) and taking the limit $\tau\rightarrow\infty$
while keeping $v$ or $a$ constant, the leading order contribution
of $O(\tau^{0})$ in Eq. (\ref{eq:EMT01}) gives \begin{eqnarray}
0 & = & (1-a^{2})\frac{d^{2}f_{p}}{dz^{2}}+f_{p}\frac{(1+a)^{2}}{1-a}\omega^{2}\cosh^{2}y-f_{p}(1-a)(-\omega\sinh y+p_{3}\cosh y)^{2}.\end{eqnarray}
We consider the central rapidity $y=0$ and static limit $p_{3}=0$,
the above equation becomes \begin{equation}
\frac{d^{2}f_{p}}{dz^{2}}+\frac{(1+a)}{(1-a)^{2}}\omega^{2}f_{p}=0.\label{eq:eom-f}\end{equation}
The reason for choosing central rapidity is that the metric (\ref{eq:dim11-metric})
can only be treated as an extension of the standard AdS one at central
rapidity, see Appendix \ref{sec:app-b}. Considering the solution
near the horizon at $a_{H}=1$ and changing variables, \begin{eqnarray}
z & \rightarrow & z'=\left(\frac{\rho_{0}}{3}\right)^{1/4}\tau^{-1/3}z,\nonumber \\
\omega' & \rightarrow & \frac{1}{2\sqrt{2}}\left(\frac{\rho_{0}}{3}\right)^{-1/4}\tau^{1/3}\omega,\label{eq:change-var}\end{eqnarray}
Eq. (\ref{eq:eom-f}) can be rewritten\begin{equation}
\frac{d^{2}f_{p}}{d{z'}^{2}}+\frac{1}{(1-z')^{2}}{\omega'}^{2}f_{p}=0.\label{eq:eom-f1}\end{equation}
The solution to Eq. (\ref{eq:eom-f1}) can be found, \begin{equation}
f_{p}(z)=(1-z')^{1/2(1\pm\sqrt{1-4{\omega'}^{2}})}\approx(1-z')^{1/2\pm i\omega'},\label{eq:solution-f}\end{equation}
where we have approximated $\sqrt{1-4{\omega'}^{2}}\approx i2\omega'$
at the limit $\tau\rightarrow\infty$, since $\omega'\sim\tau^{1/3}\gg1$.
Using the incoming wave solution corresponding to the positive sign
in (\ref{eq:solution-f}) and substituting it back into the action
(\ref{eq:action01-1}), the boundary term at $z=z_{H}$ or the black
hole horizon at $a_{H}=1$, the retarded Green function $G_{12,12}^{R}(\omega)$
at the static limt can be obtained, which is vanishing due to the
presence of the real part $\frac{1}{2}$ in the exponent of the solution
(\ref{eq:solution-f}). This can be seen from the fact that there
is a factor $(1-z')$ which is zero at the boundary, see Eq. (\ref{eq:sol1})
in Appendix (\ref{sec:app-a}). Therefore we have shown that the shear
viscosity is absent from the scaling solution in 1+1 dimension in
the leading order. 

Picking up the next-to-leading order contribution of $O(1/\tau^{1/3})$
in Eq. (\ref{eq:EOM001}), Eq. (\ref{eq:eom-f}) becomes \begin{equation}
\frac{d^{2}f_{p}}{dz^{2}}-\frac{3+5a^{2}}{z(1-a^{2})}\frac{df_{p}}{dz}+\frac{(1+a)}{(1-a)^{2}}\omega^{2}f_{p}=0.\label{eq:eom-f2}\end{equation}
Changing variables as in Eq. (\ref{eq:change-var}), the above equation
near the horizon can be rewritten in the form \begin{equation}
\frac{d^{2}f_{p}}{d{z'}^{2}}-\frac{1}{1-z'}\frac{df_{p}}{dz'}+\frac{1}{(1-z')^{2}}{\omega'}^{2}f_{p}=0,\label{eq:eom-f3}\end{equation}
whose solutions are \begin{equation}
f_{\pm p}(z)=(1-z')^{\pm i\omega'}.\label{eq:solution-f1}\end{equation}
As derived in Eq. (\ref{eq:sol2}) in Appendix \ref{sec:app-a}, the
retarded Green function at static limit is obtained, \begin{equation}
G_{12,12}^{R}(\omega,0)=-2F(\omega)=-i\frac{\sqrt{2}}{3^{3/4}\times4\pi^{2}}N^{2}\rho_{0}^{3/4}\tau^{-1}\omega.\end{equation}
The shear viscosity per unit transverse area can be obtained from
the Kubo formula (\ref{eq:kubo01}), \begin{equation}
\eta=\frac{\rho_{0}^{3/4}\sqrt{N}}{6^{3/4}\sqrt{\pi}}\frac{1}{\tau}.\label{eq:shear01}\end{equation}
where we have recovered the factor $\rho_{0}\rightarrow\rho_{0}\left(\frac{N^{2}}{2\pi^{2}}\right)^{-1}$. 

The Hawking temperature is $T=\frac{\sqrt{2}}{\pi z_{H}}$ from Eq.
(\ref{Hawking temp}) where $z_{H}$ is the horizon of the black hole
given in Eq. (\ref{eq:bh-zh}). Then the initial energy density is
a constant due to $T\sim\tau^{-1/3}$, \begin{equation}
T=\frac{\sqrt{2}}{\pi z_{H}}=\frac{\sqrt{2}}{\pi}\left(\frac{2\pi^{2}\rho_{0}}{3N^{2}}\right)^{1/4}\tau^{-1/3}.\end{equation}
The entropy per unit rapidity and unit transverse area is given by
\citep{Janik:2005zt,Sin:2006pv},\begin{equation}
S=\left(\frac{N^{2}}{2\pi}\right)^{1/4}\left(\frac{\pi}{3}\right)^{3/4}2\sqrt{2}\rho_{0}^{3/4}=\frac{N^{2}}{2}\pi^{2}T^{3}\tau.\end{equation}
The entropy density is obtained, \begin{equation}
s=\frac{S}{\tau}=\frac{1}{2}N_{c}^{2}\pi^{2}T^{3},\label{eq:entropy01}\end{equation}
where we see that the entropy density has an asymptotic behavior $s\sim\tau^{-1}$.
From Eq. (\ref{eq:shear01}) and (\ref{eq:entropy01}), we get the
well-known value, \begin{equation}
\frac{\eta}{s}=\frac{1}{4\pi}.\end{equation}

\subsection{2+1 dimension with transverse expansion \label{sec:ratio2+1}}

The metric in 2+1 dimension with radial flow is given in (\ref{eq:solution01}),
in order to calculate the shear viscosity we have to introduce a perturbation
to the background metric. It is then convenient to explicitly use
rectangular transverse coordinates $(x_{1},x_{2})$ instead of cylindrical
ones $(r,\theta)$. The corresponding line element is \begin{equation}
ds^{2}=\frac{1}{z^{2}}\left\{ -\frac{(1-a)^{2}}{1+a}dt^{2}+(1+a)(dx_{3}dx_{3}+dx_{i}dx_{i})+\frac{8}{3}\alpha\frac{(1-a)^{2}}{1+a}\frac{1}{r}x_{i}dx_{i}dt+\frac{dz^{2}}{z^{2}}\right\} ,\label{eq:back-tr}\end{equation}
where $r=\sqrt{x_{i}x_{i}}$ and the summation over $i=1,2$ is implied.
We keep in mind that $\alpha$ is small. The metric determinant has
the same form as in the 1+1 dimensional case in the leading order,
\begin{equation}
\sqrt{-g}=\frac{1}{z^{5}}(1-a^{2})+O(\alpha^{2}).\end{equation}
Now we consider the perturbation $h_{12}(t,x_{3},z)$ to the background
metric (\ref{eq:back-tr}). Denoting $\phi=h_{2}^{1}(t,x_{3},z)$,
we find the equation of motion for $\phi$, \begin{equation}
0=\delta\mathcal{R}_{2}^{1}+4\phi=-\frac{1}{2\sqrt{-g}}\partial_{\mu}(\sqrt{-g}g^{\mu\nu}\partial_{\nu}\phi).\label{eq:eom-h12}\end{equation}
It can be verified that the entangled term $\sim\alpha\phi$ appear
in $O(t^{-1/3})$, while leading terms are of $O(t^{2/3})$. So the
transverse part is decoupled from $\phi$ in the leading order. The
shear viscosity and entropy density per unit transverse area are the
same as in the 1+1 dimensional case, Eqs. (\ref{eq:shear01}) and
(\ref{eq:entropy01}). 

We can also consider the perturbation along $x_{2}$ axis (or equivalently
$x_{1}$ axis), i.e. $\phi=h_{3}^{1}(t,z,x_{2})$. To the leading
order $O(\tau^{2/3})$ we find the equation of motion for $\phi$,
\begin{equation}
0=\delta\mathcal{R}_{3}^{1}+4\phi=-\frac{1}{2\sqrt{-g}}\partial_{\mu}(\sqrt{-g}g^{\mu\nu}\partial_{\nu}\phi)+t^{2/3}\frac{2}{3}\alpha\frac{1}{1+a}\frac{(x^{2})^{2}}{r^{3}}\partial_{t}\phi,\label{eq:eom-h13-01}\end{equation}
whose explicit form reads, \begin{equation}
-\partial_{z}^{2}\phi+\frac{1+a}{(1-a)^{2}}\partial_{t}^{2}\phi-\frac{4}{3}\alpha\frac{1}{1+a}\frac{(x^{1})^{2}}{r^{3}}\partial_{t}\phi=0.\label{eq:eom-h13-02}\end{equation}
We see that the transverse part enters the equation of motion in the
leading order. We assume that the solution has the factorization form
$\phi(t,x_{2},z)=\phi_{0}(p)f_{p}(z)e^{-i\omega t+ip_{2}x_{2}}$,
then we derive from Eq. (\ref{eq:eom-h13-02}) the differential equation
for $f_{p}(z)$,

\begin{equation}
\frac{d^{2}f_{p}}{dz^{2}}-\frac{3+5a}{z(1-a^{2})}\frac{df_{p}}{dz}+\frac{1+a}{(1-a)^{2}}\omega^{2}f_{p}-\frac{i}{1+a}\frac{2\alpha}{3r}\omega f_{p}=0,\end{equation}
where we can expand the equation near the horzion $z\rightarrow z_{H}$
in Eq. (\ref{eq:bh-zh}), 

\begin{equation}
\frac{d^{2}f_{p}}{d{z'}^{2}}-\frac{1}{1-z'}\frac{df_{p}}{dz'}+\frac{1}{(1-z')^{2}}{\omega'}^{2}f_{p}-i\alpha\frac{2\sqrt{2}z_{H}}{3r}\omega'f_{p}=0.\label{eq:EOM-21}\end{equation}
The solution is a linear combination of Bessel functions, \begin{eqnarray*}
f_{p1} & = & J_{i\omega'}\left[i^{3/2}(1-z')z_{H}\sqrt{\frac{\alpha\omega}{3r}}\right],\\
f_{p2} & = & Y_{i\omega'}\left[i^{3/2}(1-z')z_{H}\sqrt{\frac{\alpha\omega}{3r}}\right].\end{eqnarray*}
Since $\alpha$ is small, we can make expansion in $\alpha$ for the
solutions $f_{p1}$ and $f_{p2}$. One can verify that the solution
is a linear combination of $f_{p}$ and $f_{-p}$, \begin{eqnarray}
f_{p} & = & (1-z')^{i\omega'}z_{H}^{i\omega'}\left(\frac{\alpha\omega}{3r}\right)^{i\omega'}\left[\frac{(-1)^{i3\omega'/4}2^{-i\omega'}}{\Gamma(1+i\omega')}+\frac{i(-1)^{i3\omega'/4}2^{-2-i\omega'}}{(1+i\omega')\Gamma(1+i\omega')}(1-z')^{2}z_{H}^{2}\frac{\alpha\omega}{3r}\right],\nonumber \\
f_{-p} & = & f_{p}(i\omega'\rightarrow-i\omega').\label{eq:sol21}\end{eqnarray}
One sees that $f_{-p}=f_{p}^{*}$. Following Eq. (\ref{eq:sol3})
in Appendix \ref{sec:app-a} and steps in previous section, we get
the same value as in the case of 1+1 dimension, i.e. $\eta/s=1/(4\pi)$.

\section{Summary and discussions \label{sec:Conclusion}}

We derive a time dependent metric dual to sQGP fluid in 2+1 dimension
with radial flow in late time by holographic renormalization. It is
difficult to obtain the exact solution to the Einstein equation with
this metric, especially when the metric has off-diagonal components
for radial flows. If transverse expansion is small and can be treated
as a perturbation, the late time asymptotic solution, the metric in
(\ref{eq:solution01}) with off-diagonal elements, can be found by
using $v=z^{4}/\tau^{4/3}$ as a scaling and expansion parameter.
With this metric we calculate the ratio $\eta/s$ of shear viscosity
$\eta$ to entropy density $s$ for sQGP in SYM field theory with
the KSS method. As a first attempt we consider 1+1 dimension with
only longitudinal flow whose metric is diagonal. If we include only
the leading order terms in the equation of motion for perturbations
to the metric, the shear viscosity is vanishing, consistent to the
assumption that the fluid is ideal. We reproduce KSS bound $1/(4\pi)$
for $\eta/s$ if we pick up the next-to-leading order term in the
equation of motion, indicating that the shear viscosity is a higher
order effect. Our derivation is based on the Janik and Peschanski's
method and is valid in late time, $\tau\rightarrow\infty$. For intermidiate
stage of hydrodynamic evolution the ratio is not necessarily $1/(4\pi)$
for an expanding fluid, so our result is not trivial or obvious. We
further show that the ratio for fluids in 2+1 dimension in late time
with transverse flow is the same as in 1+1 dimension in the leading
order of transverse rapidity. We remember that the mean free path
is $l_{\mathrm{mfp}}\sim1/(n\sigma v)$, where $n\sim T^{3}$ is the
particle number density, $\sigma\sim g^{2}T^{-2}$ ($g$ is the coupling
constant) the typical scattering cross section and $v\sim1$ the typical
velocity. The shear viscosity can then be estimated as $\eta\sim\rho l_{\mathrm{mfp}}\sim g^{-2}T^{3}$,
where $\rho$ is the energy density. In comparsion with our result
in Eq. (\ref{eq:shear01}), we have $\eta\sim\tau^{-1}$, $T\sim\tau^{-1/3}$
and $g$ does not change with the proper time. This implies that if
the coupling is as strong at the beginning as in late time of fluid
expansion or hydrodynamic evolution does not influence the strength
of the interaction. 

In 1+1 dimension one can introduce the shear viscosity of the next-to-leading
order in the stress tensor in late time solution \citet{Janik:2006ft,Nakamura:2006ih,Sin:2006pv}.
So an interesting attempt is to calculate the shear viscosity with
the metric dual to the stress tensor with shear terms. We found that
the correction to the shear viscosity is also of the next-to-leading
order, $\eta/s=1/(4\pi)+O(\tau^{-1-s})$. 

The same analysis can also be applied to hydrodynamic behaviors of
sQGP in early time, which is important to understand the initial state
of QGP \citet{Kovchegov:2007pq}. However early time behaviors of
fluids show anisotropic feature and therefore more complicated. We
will reserve it for a future investigation. 

\begin{acknowledgments}
We thank D. Rischke, H.-c. Ren and P.-f. Zhuang for helpful discussions.
Q.W. is supported in part by '100 talents' project of Chinese Academy
of Sciences (CAS), by National Natural Science Foundation of China
(NSFC) under the grants 10675109 and 10735040. 
\end{acknowledgments}
\appendix

\section{Derivation of the metric in (\ref{eq:static-metric})}

\label{sec:app-b}The D3 black AdS metric can be writen as \begin{equation}
ds^{2}=-\frac{(1-z^{4}/z_{0}^{4})^{2}}{1+z^{4}/z_{0}^{4}}dt^{2}+(1+z^{4}/z_{0}^{4})\frac{d\mathbf{x}^{2}}{z^{2}}+\frac{dz^{2}}{z^{2}}.\label{eq:static-metric01}\end{equation}
Changing variables to $(t,x^{3})\rightarrow(\tau,y)$ through $t=\tau\cosh y$,
$x^{3}=\tau\sinh y$, the metric becomes \begin{eqnarray}
ds^{2} & = & (-A\cosh^{2}y+B\sinh^{2}y)d\tau^{2}+\tau^{2}(-A\sinh^{2}y+B\cosh^{2}y)dy^{2}\nonumber \\
 &  & +2\tau(-A+B)\cosh y\sinh yd\tau dy+...\label{eq:temp-metric01}\end{eqnarray}
where $A=\frac{(1-z^{4}/z_{0}^{4})^{2}}{1+z^{4}/z_{0}^{4}}$ and $B=(1+z^{4}/z_{0}^{4})$.
The metric (\ref{eq:temp-metric01}) has the symmetry under $y\rightarrow-y$.
In comparsion with Janik and Peschanski's time dependence metric,
there is an off-diagonal part $\sim d\tau dy$ in the metric (\ref{eq:temp-metric01}).
So the metric (\ref{eq:static-metric}) is only valid in central rapidity
region around $y\sim0$.

\section{Evaluation of retarded Green function from solution (\ref{eq:solution-f})}

\label{sec:app-a}Substituting the solutions $f_{\pm}$ in (\ref{eq:solution-f})
and (\ref{eq:solution-f1}) back into the action (\ref{eq:action01})
and keep the boundary term at $z=z_{H}$, we find 

\begin{eqnarray*}
I_{5D} & = & -\frac{N^{2}}{16\pi^{2}}\int d^{4}xdz\frac{\partial}{\partial z}[\sqrt{-g}g^{zz}\phi(\tau,y,z)\partial_{z}\phi(\tau,y,z)]\\
 & = & -\frac{N^{2}}{16\pi^{2}}\int d^{4}x\tau\left.\frac{1-a^{2}}{z^{3}}\phi(\tau,y,z)\partial_{z}\phi(\tau,y,z)\right|_{z=z_{H}}\\
 & = & -\frac{N^{2}}{16\pi^{2}}\int d\tau dydx^{1}dx^{2}\tau\int\frac{d\omega dp_{3}d\omega'dp_{3}'}{(2\pi)^{4}}\exp[-i(\omega+\omega')\tau\cosh y+i(p_{3}+p_{3}')\tau\sinh y]\\
 &  & \times\phi_{0}(\omega,p_{3})\phi_{0}(\omega',p_{3}')\left.\frac{1-a^{2}}{z^{3}}f_{p'}(z)\frac{df_{p}(z)}{dz}\right|_{z=z_{H}}\\
 & = & -\frac{N^{2}}{16\pi^{2}}\int d(\tau\cosh y)d(\tau\sinh y)dx^{1}dx^{2}\int\frac{d\omega dp_{3}d\omega'dp_{3}'}{(2\pi)^{4}}\\
 &  & \times\exp[-i(\omega+\omega')\tau\cosh y+i(p_{3}+p_{3}')\tau\sinh y]\phi_{0}(\omega',p_{3}')\phi_{0}(\omega,p_{3})\\
 &  & \times\left.\frac{1-a^{2}}{z^{3}}f_{p'}(z)\frac{df_{p}(z)}{dz}\right|_{z=z_{H}}\\
 & = & -\frac{N^{2}}{16\pi^{2}}\int dx^{1}dx^{2}\int\frac{d\omega dp_{3}}{(2\pi)^{2}}\phi_{0}(-\omega,-p_{3})\phi_{0}(\omega,p_{3})\left.\frac{1-a^{2}}{z^{3}}f_{-p}(z)\frac{df_{p}(z)}{dz}\right|_{z=z_{H}}\\
 & \equiv & \int\frac{d^{4}p}{(2\pi)^{4}}\phi_{0}(-\omega,-p)\phi_{0}(\omega,p)F(\omega).\end{eqnarray*}
In the fourth equality we have changed the integral variables, $d\tau dy\tau=d(\tau\cosh y)d(\tau\sinh y)$.
In the last we have assumed that $\phi_{0}(\omega,p)=(2\pi)^{2}\delta(p_{1})\delta(p_{2})\phi_{0}(-\omega,-p_{3})$
and set the transverse area $L^{2}=1$. Now we evaluate $F(\omega)$.
For the solutions (\ref{eq:solution-f}), we have \begin{eqnarray}
F(\omega) & \propto & \lim_{z\rightarrow z_{H}}\frac{1-a^{2}}{z^{3}}f_{-p}(z)\frac{df_{p}(z)}{dz}\nonumber \\
 & = & \frac{8}{z_{H}^{3}}\lim_{z\rightarrow z_{H}}(1-z')(1-z')^{1/2-i\omega'}\frac{d}{dz}(1-z')^{1/2+i\omega'}\nonumber \\
 & = & -\frac{8}{3}\rho_{0}\tau^{-4/3}(1/2+i\omega')\lim_{z\rightarrow z_{H}}(1-z')\nonumber \\
 & = & 0,\label{eq:sol1}\end{eqnarray}
while for the solutions (\ref{eq:solution-f1}), we obtain \begin{eqnarray}
F(\omega) & = & -\frac{N^{2}}{16\pi^{2}}\left.\frac{1-a^{2}}{z^{3}}f_{-p}(z)\frac{df_{p}(z)}{dz}\right|_{z=z_{H}}\nonumber \\
 & = & i\frac{N^{2}}{16\pi^{2}}\omega'\frac{8}{3}\rho_{0}\tau^{-4/3}\lim_{z\rightarrow z_{H}}(1-z')(1-z')^{-i\omega'}(1-z')^{i\omega'-1}\nonumber \\
 & = & i\omega\frac{\sqrt{2}N^{2}}{8\pi^{2}}\left(\frac{\rho_{0}}{3}\right)^{3/4}\tau^{-1}.\label{eq:sol2}\end{eqnarray}
For the solution (\ref{eq:sol21}), we get \begin{eqnarray}
F(\omega) & = & -\frac{N^{2}}{16\pi^{2}}\left.\frac{1-a^{2}}{z^{3}}f_{-p}(z)\frac{df_{p}(z)}{dz}\right|_{z=z_{H}}\nonumber \\
 & = & -\frac{N^{2}}{16\pi^{2}}\frac{8}{3}\rho_{0}\tau^{-4/3}\lim_{z\rightarrow z_{H}}(1-z')f_{-p}(z)\frac{df_{p}(z)}{dz'}\nonumber \\
 & = & i\omega\frac{\sqrt{2}N^{2}}{8\pi^{2}}\left(\frac{\rho_{0}}{3}\right)^{3/4}\tau^{-1}\label{eq:sol3}\end{eqnarray}
where we have used \begin{eqnarray*}
\frac{df_{p}(z)}{dz'} & = & -i\omega'(1-z')^{i\omega'-1}z_{H}^{i\omega'}\left(\frac{\alpha\omega}{3r}\right)^{i\omega'}\left[\frac{(-1)^{i3\omega'/4}2^{-i\omega'}}{\Gamma(1+i\omega')}+\frac{i(-1)^{i3\omega'/4}2^{-2-i\omega'}}{(1+i\omega')\Gamma(1+i\omega')}(1-z')^{2}z_{H}^{2}\frac{\alpha\omega}{3r}\right]\\
 &  & -2(1-z')^{i\omega'+1}z_{H}^{i\omega'+2}\left(\frac{\alpha\omega}{3r}\right)^{i\omega'+1}\frac{i(-1)^{i3\omega'/4}2^{-2-i\omega'}}{(1+i\omega')\Gamma(1+i\omega')}\end{eqnarray*}


\begin{thebibliography}{10}
\bibitem{Lee:1974ma}T.~D.~Lee and G.~C.~Wick, Phys.\ Rev.\ D
{\bf 9}, 2291 (1974). 

\bibitem{Karsch:2000ps}F.~Karsch, E.~Laermann and A.~Peikert,
Phys.\ Lett.\ B {\bf 478}, 447 (2000) {[}arXiv:hep-lat/0002003].

\bibitem{Hofmann:1974}J.~Hofmann, H.~Stocker, W.~Scheid and W.~Greiner,
Bear Mountain Workshop, New York, Dec 1974. 

\bibitem{gyulassy:2005}M.~Gyulassy, L.~McLerran, Nucl.\ Phys.\ A{\bf 750},
30(2005). 

\bibitem{Shuryak:2004cy}E.~V.~Shuryak, Nucl.\ Phys.\ A {\bf 750},
64 (2005) {[}arXiv:hep-ph/0405066]. 

\bibitem{Maldacena:1997re}J.~M.~Maldacena, Adv.\ Theor.\ Math.\ Phys.\ {\bf 2},
231 (1998) {[}Int.\ J.\ Theor.\ Phys.\ {\bf 38}, 1113 (1999)]
{[}arXiv:hep-th/9711200]. 

\bibitem{Gubser:1998bc}S.~S.~Gubser, I.~R.~Klebanov and A.~M.~Polyakov,
Phys.\ Lett.\ B {\bf 428}, 105 (1998) {[}arXiv:hep-th/9802109].

\bibitem{Witten:1998zw}E.~Witten, Adv.\ Theor.\ Math.\ Phys.\ {\bf 2},
505 (1998) {[}arXiv:hep-th/9803131]. 

\bibitem{Policastro:2001yc}G.~Policastro, D.~T.~Son and A.~O.~Starinets,
Phys.\ Rev.\ Lett.\ {\bf 87}, 081601 (2001) {[}arXiv:hep-th/0104066].

\bibitem{Kovtun:2004de}P.~Kovtun, D.~T.~Son and A.~O.~Starinets,
Phys.\ Rev.\ Lett.\ {\bf 94}, 111601 (2005) {[}arXiv:hep-th/0405231].

\bibitem{Buchel:2003tz}A.~Buchel and J.~T.~Liu, Phys.\ Rev.\ Lett.\ {\bf 93},
090602 (2004) {[}arXiv:hep-th/0311175]. 

\bibitem{Maeda:2006by}K.~Maeda, M.~Natsuume and T.~Okamura, Phys.\ Rev.\ D
{\bf 73}, 066013 (2006) {[}arXiv:hep-th/0602010]. 

\bibitem{Natsuume:2007ty}M.~Natsuume and T.~Okamura, Phys.\ Rev.\ D
{\bf 77}, 066014 (2008) {[}arXiv:0712.2916 {[}hep-th]]. 

\bibitem{Baier:2007ix}R.~Baier, P.~Romatschke, D.~T.~Son, A.~O.~Starinets
and M.~A.~Stephanov, JHEP {\bf 0804}, 100 (2008) {[}arXiv:0712.2451
{[}hep-th]]. 

\bibitem{Liu:2006ug}H.~Liu, K.~Rajagopal and U.~A.~Wiedemann,
Phys.\ Rev.\ Lett.\ {\bf 97}, 182301 (2006) {[}arXiv:hep-ph/0605178].

\bibitem{Herzog:2006gh}C.~P.~Herzog, A.~Karch, P.~Kovtun, C.~Kozcaz
and L.~G.~Yaffe, JHEP {\bf 0607}, 013 (2006) {[}arXiv:hep-th/0605158].

\bibitem{Gubser:2006bz}S.~S.~Gubser, Phys.\ Rev.\ D {\bf 74},
126005 (2006) {[}arXiv:hep-th/0605182]. 

\bibitem{Liu:2006nn}H.~Liu, K.~Rajagopal and U.~A.~Wiedemann,
Phys.\ Rev.\ Lett.\ {\bf 98}, 182301 (2007) {[}arXiv:hep-ph/0607062].

\bibitem{Peeters:2006iu}K.~Peeters, J.~Sonnenschein and M.~Zamaklar,
Phys.\ Rev.\ D {\bf 74}, 106008 (2006) {[}arXiv:hep-th/0606195].

\bibitem{Hou:2007uk}D.~Hou and H.~c.~Ren, JHEP {\bf 0801}, 029
(2008) {[}arXiv:0710.2639 {[}hep-ph]]. 

\bibitem{Li:2008py}M.~Li, Y.~Zhou and P.~Pu, JHEP {\bf 0810},
010 (2008) {[}arXiv:0805.1611 {[}hep-th]]. 

\bibitem{Song:2007ux} H.~Song and U.~W.~Heinz, arXiv:0712.3715
{[}nucl-th]. 

\bibitem{Chaudhuri:2008sj} A.~K.~Chaudhuri,  arXiv:0801.3180 {[}nucl-th].

\bibitem{Bjorken:1982qr} J.~D.~Bjorken, Phys.\ Rev.\ D {\bf 27},
140 (1983). 

\bibitem{Janik:2005zt}R.~A.~Janik and R.~Peschanski, Phys.\ Rev.\ D
\textbf{73}, 045013 (2006) {[}arXiv:hep-th/0512162]. 

\bibitem{Janik:2006ft}R.~A.~Janik, Phys.\ Rev.\ Lett.\ \textbf{98},
022302 (2007) {[}arXiv:hep-th/0610144]. 

\bibitem{Benincasa:2007tp}P.~Benincasa, A.~Buchel, M.~P.~Heller
and R.~A.~Janik, Phys.\ Rev.\ D {\bf 77}, 046006 (2008) {[}arXiv:0712.2025
{[}hep-th]]. 

\bibitem{Nakamura:2006ih}S.~Nakamura and S.~J.~Sin, JHEP \textbf{0609},
020 (2006) {[}arXiv:hep-th/0607123] 

\bibitem{Sin:2006pv}S.~J.~Sin, S.~Nakamura and S.~P.~Kim, JHEP
\textbf{0612}, 075 (2006) {[}arXiv:hep-th/0610113].  

\bibitem{Kovchegov:2007pq}Y.~V.~Kovchegov and A.~Taliotis, Phys.\ Rev.\ C
{\bf 76}, 014905 (2007) {[}arXiv:0705.1234 {[}hep-ph]]. 

\bibitem{Albacete:2008vs}J.~L.~Albacete, Y.~V.~Kovchegov and
A.~Taliotis, JHEP {\bf 0807}, 100 (2008) {[}arXiv:0805.2927 {[}hep-th]].

\bibitem{Kajantie:2007bn}K.~Kajantie, J.~Louko and T.~Tahkokallio,
Phys.\ Rev.\ D {\bf 76}, 106006 (2007) {[}arXiv:0705.1791 {[}hep-th]].

\bibitem{Son:2002sd}D.~T.~Son and A.~O.~Starinets, JHEP \textbf{0209},
042 (2002) {[}arXiv:hep-th/0205051]. 

\bibitem{Son:2007vk}D.~T.~Son and A.~O.~Starinets, Ann.\ Rev.\ Nucl.\ Part.\ Sci.\ \textbf{57},
95 (2007) {[}arXiv:0704.0240 {[}hep-th]]. 

\bibitem{Kolb:2003dz}P.~F.~Kolb and U.~W.~Heinz, arXiv:nucl-th/0305084.

\bibitem{fefferman1985}D.~Z.~Fefferman and C.~R.~Graham, Conformal
invariants, Elie Cartan et les Mathématiques d'aujourd'hui (Astérisque,
1985).

\bibitem{Karch:2005ms}A.~Karch, A.~O'Bannon and K.~Skenderis,
JHEP \textbf{0604}, 015 (2006) {[}arXiv:hep-th/0512125]. 

\bibitem{Skenderis:2002wp}K.~Skenderis, Class.\ Quant.\ Grav.\ \textbf{19},
5849 (2002) {[}arXiv:hep-th/0209067]. 
\end{thebibliography}
\end{document}